\documentclass{aastex}
\usepackage{emulateapj5,epsfig}

\newenvironment{inlinefigure}{ 
\def\@captype{figure} 
\noindent\begin{minipage}{0.999\linewidth}\begin{center}} 
{\end{center}\end{minipage}\smallskip} 
\newcommand{\beq}{\begin{equation}}
\newcommand{\eeq}{\end{equation}}

\usepackage{graphicx}
\usepackage{amssymb}
\usepackage{epstopdf}
\font\tenbg=cmmib10 at 10pt

\def \rvecphi{{\hbox{\tenbg\char'036}}}

\def \trho{{\tilde{\rho}}}
\def\lesssim{\mathrel{\hbox{\rlap{\hbox{\lower4pt\hbox{$\sim$}}}\hbox{$<$}}}}
\def\gtrsim{\mathrel{\hbox{\rlap{\hbox{\lower4pt\hbox{$\sim$}}}\hbox{$>$}}}}

\begin{document}
\title{Advection/Diffusion of Large-Scale B-Field
in Accretion Disks}

\author{
 R.V.E. Lovelace\altaffilmark{1},  D.M. Rothstein\altaffilmark{2},
 \& G.S. Bisnovatyi-Kogan\altaffilmark{3}}

\altaffiltext{1}{Department of Astronomy
and Department of Applied and Engineering
Physics,
Cornell University, Ithaca, NY 14853-6801; 
RVL1@cornell.edu}

\altaffiltext{2}{Department of   Astronomy
Cornell University, Ithaca, NY 14853-6801; 
droth@astro.cornell.edu}

\altaffiltext{3}{Space Research Institute,
Russian Academy of Sciences, Moscow, Russia;
gkogan@mx.iki.rssi.ru}

\begin{abstract}

  Activity of the nuclei of galaxies and stellar
mass systems involving disk accretion to
black holes is thought to be due to (1)
a small-scale turbulent magnetic field
in the disk (due to the magneto-rotational
instability or MRI) which gives a large
viscosity enhancing accretion, and (2)
a large-scale magnetic field which gives
rise to matter outflows and/or electromagnetic
jets from the disk which also enhances 
accretion.   
   An important problem with this picture
is that the enhanced viscosity is accompanied by
an enhanced magnetic diffusivity which acts to
prevent the build up of a significant large-scale field.
    Recent work has
pointed out that   the disk's surface
layers  are non-turbulent and
thus highly conducting (or non-diffusive)
because the MRI is suppressed 
high in the disk  where    the magnetic 
and radiation pressures
are  larger than the thermal pressure.
     Here, we calculate the vertical
($z$)      profiles 
of the stationary accretion flows (with radial
and azimuthal components), and
the profiles of the  large-scale,
magnetic field taking into
account the turbulent viscosity and diffusivity
due to the MRI and the fact that the turbulence 
vanishes at the surface of the disk.   
   We derive a sixth-order differential equation
for the radial flow velocity $v_r(z)$ which 
depends mainly on the midplane
thermal to magnetic pressure ratio $\beta >1$ and
the Prandtl number of the turbulence
${\cal P}=$ viscosity/diffusivity. 
     Boundary conditions at the disk
surface take into account a possible magnetic
wind or jet and allow for a surface current
in the highly conducting surface layer.
   The stationary solutions we find indicate that a
weak ($\beta > 1$) large-scale field  does not diffuse away as suggested
by earlier work.
     For a wide range of parameters $\beta >1 $ and ${\cal  P}\geq 1$,
we find stationary {\it channel} type flows where 
the  flow is radially {\it outward} near the midplane
of the disk and  radially {\it inward} in the top and bottom parts
of the disk.     Channel flows with inward flow near the midplane
and outflow in the top and bottom parts of the disk are also found.
    We find that  Prandtl numbers larger than a critical value 
(estimated to be $2.7$)
are needed in order for there to be magnetocentrifugal outflows from
the disk's surface.  For smaller ${\cal P}$, electromagnetic outflows are
predicted.

\end{abstract}

\keywords{accretion, accretion disks --- galaxies: jets --- magnetic
fields --- MHD --- X-rays: binaries
}

\section{Introduction}

    Early work on disk accretion to a black hole argued that a
large-scale poloidal magnetic field originating from say the interstellar medium,
would be dragged inward and greatly compressed near
the black hole by the accreting plasma (Bisnovatyi-Kogan \& Ruzmaikin 1974, 1976)
and that this would be important for the formation of jets
(Lovelace 1976). 
      Later, the importance of a weak 
small-scale magnetic field within the disk was
recognized as the source of the turbulent viscosity of
disk owing to the magneto-rotational instability
(MRI; Balbus \& Hawley 1991, 1998).
         Analysis of the diffusion and advection of a
large-scale field in a disk with a turbulent
viscosity comparable to the turbulent magnetic
diffusivity (as suggested by MRI simulations) 
indicated that a {\it weak} large-scale field would
diffuse outward rapidly (van Ballegooijen 1989; 
Lubow, Papaloizou, \& Pringle 1994;
Lovelace, Romanova, \& Newman 1994, 1997).
     This has led to the suggestion that special conditions
(nonaxisymmetry) are required for the field to be advected
inward (Spruit \& Uzdensky 2005).
       Recently,  Bisnovatyi-Kogan and Lovelace (2007) 
pointed out that  the disk's surface
layers are highly conducting (or non-diffusive)
because  the MRI is suppressed
in this region where    the
magnetic energy-density is larger than
the thermal energy-density.
   Rothstein and Lovelace (2008)  analyzed this problem
in further detail and discussed the connections
with global and shearing box magnetohydrodynamic
(MHD) simulations of the MRI.   

      With the disk's highly conducting surface
neglected, the  fast outward diffusive drift of the
large-scale poloidal magnetic
field in a turbulent disk can be readily understood by looking at the induction
equation for the vertical field  $B_z$,
$$
{\partial (rB_z) \over \partial t} =
{\partial \over \partial r}\left[  (r B_z)u - \eta r 
\left( {\partial B_r \over \partial z}-{\partial B_z \over \partial r }\right)\right]~,   
$$ 
where $u >0$ is the accretion speed and $\eta$ is the turbulent diffusivity
both assumed independent of $z$ (Lovelace et al. 1994). 
(Both assumptions are invalid and are removed in this work.)
   For a dipole type field,  $B_z$ is  an even function of 
$z$ and it changes very little over the  half-thickness $h$ of a thin disk
($h\ll r$).  
   Consequently, the average of this equation over the half-thickness  gives 
$$
{\partial (rB_z) \over \partial t} =
{\partial \over \partial r}\left[ (r B_z)u  - \left({ r\over h} \right)\eta(B_r)_h
+\eta r {\partial B_z \over \partial r}\right]~,
$$ 
where $(B_r)_h$ is the radial field at the disk's surface.
   The terms inside the square brackets represents the flux of the
$z-$magnetic flux.  For stationary conditions it vanishes.
    The term proportional to $u$ describes
 the {\it inward advection} of the magnetic field, the term $\propto \eta (B_r)_h$ describes the {\it outward diffusive drift} of the field, and the last term represents the {\it radial diffusion} of the field.   For $(B_r)_h \sim B_z$ the
 radial diffusion term is negligible compared with the diffusive drift term.
 For a weak $B_z$ field and the standard $\alpha-$disk model with turbulent viscosity of the order of the turbulent diffusivity, the accretion speed is
 $u \sim \alpha c_s h/r$ (with $c_s$ the sound speed), whereas
 the outward drift speed of the field is $ \sim \alpha c_s$ which is
 a factor $r/h \gg 1$ larger than $u$.   
 That is, a stationary solution with $B_z\neq 0$ is not possible. 
   A stationary or growing $B_z$ field is possible if the field
is sufficiently strong to give appreciable outflow
of angular momentum to jets (Lovelace et al. 1994).   
 
       Three-dimensional MHD simulations
have been performed which give some information
on the advection/diffusion of a large scale field.       
These simulations resolve the largest 
scales of the MRI
turbulence and therefore self-consistently include
the turbulent viscosity and diffusivity. 
Most simulations performed to date have
investigated conditions in which the accreting matter 
contains no net magnetic flux and where no magnetic field is supplied
at the boundary of the computational domain 
(e.g.,   Hirose et al. 2004; De Villiers et al. 2005; Hawley
\& Krolik 2006;  McKinney \& Narayan 2007).
   In these simulations
stretching of locally poloidal field lines in the initial configuration
leads to a large-scale poloidal fields and jet structures in the
inner disk.
    However, in simulations by Igumenshchev, Narayan, \& Abramowicz 2003;
Igumenshchev 2008)
weak poloidal flux injected at the outer boundary
is clearly observed to be dragged into the central region of the
disk, leading to the buildup of a strong poloidal magnetic field
close to the central object.  
    This flux build up and its limit are  discussed
by Narayan, Igumenshchev, \& Abramowicz (2003).
   The extent to which  the magnetic field advection
seen in numerical simulations depends on having
a thick disk or nonaxisymmetric conditions is unclear.

     In this work we calculate the profiles through the disk
of stationary accretion flows (with radial
and azimuthal components), and
the profiles of a  large-scale,
weak magnetic field taking into
account the turbulent viscosity and diffusivity
due to the MRI and the fact that the turbulence 
vanishes at the surface of the disk.   
    By a weak field we mean that the
magnetic pressure in the middle of the disk
is less than the thermal pressure.    
     Related calculations of the disk structure
were done earlier by K\"onigl (1989), Li (1995),  Ogilvie
and Livio (2001) but without taking
into account the  absence of turbulence
at the disk's surface.
      Recent work calls into question the $\alpha$-description
of the MRI turbulence in accretion disks and develops
a closure model which fits shearing box simulation
results (Pessah, Chan, \& Psaltis 2008).    
   Analysis of this more complicated model is deferred 
a future study.

    Section 2 develops the equations
for the flow and magnetic field in
a viscous diffusive disk.   Section 3
discusses the boundary conditions at
the surface of the disk.   Section 4 derives
the internal flow/field solutions
for an analytically soluble disk model.
  Section 5 discusses the external flow/field
solutions which may be magnetocentrifugal
winds or electromagnetic outflows.
  Section 6 gives the conclusions of this
work.

\section{Theory}

       We consider the non-ideal magnetohydrodynamics 
of a thin axisymmetric, viscous, resistive disk threaded
by a large-scale dipole-symmetry magnetic field ${\bf B}$.
  We use a cylindrical $(r,\phi,z)$ inertial coordinate system
in which the time-averaged
magnetic field is ${\bf B}=B_r\hat{\bf r}+B_\phi\hat{\rvecphi~}+
B_z\hat{\bf z}$, and the time-averaged flow velocity is
${\bf v}=v_r\hat{\bf r}+v_\phi\hat{\rvecphi~}+
v_z\hat{\bf z}$.   
    The main equations are
\begin{eqnarray}
\rho {d{\bf v}\over dt}&=&-{\bf \nabla} p +\rho{\bf g}
+ {1\over c}{\bf J \times B} + {\bf F^\nu}~,
\\
{\partial {\bf B} \over \partial t}
&=& {\nabla \times}({\bf v \times B}) - 
{\bf \nabla}\times(\eta {\bf \nabla }\times {\bf B})~.
\end{eqnarray}
These equations are supplemented  by the continuity equation,
$\nabla(\rho {\bf v})=0$, by
${\bf \nabla \times B} =4\pi{\bf J}/c$, and by
${\bf\nabla} \cdot {\bf B}=0$.
   Here, $\eta$ is the magnetic
diffusivity, ${\bf F}^\nu=-{\bf \nabla}\cdot T^\nu$ 
is the viscous force with 
$T_{jk}^\nu= -\rho \nu (\partial v_j/\partial x_k 
+\partial v_k/\partial x_j-(2/3)\delta_{jk}
{\bf \nabla} \cdot{\bf v} )$ (in Cartesian
coordinates), and $\nu$ is the
kinematic viscosity.    
     For simplicity,  in place of an energy equation
we consider the adiabatic dependence $p \propto \rho^\gamma$,
with $\gamma$ the adiabatic index.     
  
   We assume that both the viscosity and
the diffusivity are
due to  magneto-rotational (MRI) turbulence in
the disk  so that
\begin{equation}
\nu ={\cal P} \eta =\alpha ~{c_{s0}^2 \over \Omega_K}~ g(z)~,
\end{equation}
where ${\cal P}$ is the magnetic Prandtl number of
the turbulence assumed
a constant of order unity (Bisnovatyi-Kogan \& Ruzmaikin 1976),
$\alpha \leq 1$ is the  dimensionless Shakura-Sunyaev
(1973) parameter, $c_{s0}$ is the midplane isothermal sound
speed, $\Omega_K \equiv
(GM/r^3)^{1/2}$ is
the Keplerian angular velocity
of the disk, and $M$ is the
mass of the central object.
     The function  $g(z)$
accounts for the absence of turbulence in
the surface layer of the disk (Bisnovatyi-Kogan \& Lovelace
2007;  Rothstein \& Lovelace 2008).
    In the  body of the disk $g = 1$, whereas
at the surface of the disk, at say $z_S$, $g$ tends over
a short distance to
a very small value  
$\sim 10^{-8}$, effectively zero, which is the ratio of the Spitzer
diffusivity of the disk's
surface layer to the turbulent diffusivity of
the body of the disk.  
 At the disk's surface the density is much
 smaller than its midplane value.

     We consider stationary solutions of equations (1) and (2)
for a weak large-scale magnetic field.    
These can be greatly simplified
for thin disks where the disk half-thickness, of
the order of  $h \equiv c_{s0}/\Omega_K$,
is much less than $r$.
Thus we have the small parameter 
\begin{equation}
\varepsilon={h \over r} = {c_{s0}\over v_K} \ll 1~.
\end{equation}
It is useful in the following to use the
dimensionless height $\zeta \equiv z/h$.

  The three magnetic field components are  assumed
to be of comparable magnitude on the disk's surface,
but $B_r = 0 = B_\phi$ on the midplane.  
  On the other hand the axial magnetic field changes
by only a small almount going from the midplane
to the surface, $\Delta B_z \sim \varepsilon  B_r \ll B_z$
(from ${\bf \nabla} \cdot{\bf B} =0$) so that $B_z\approx$ const inside
the disk.   
   As a consequence, the 
$\partial B_j/\partial r$ terms in the 
magnetic force  in equation (1) can all be dropped
in favor of  the $\partial B_j/\partial z$ terms
(with $j=r,~\phi$).   
   Thus, we neglect the final term of the induction
equation given in the Introduction.   
   It is important to keep in mind that $B_j$ is the
large scale field;  the approximation does not
apply to the small-scale field which gives
the viscosity and diffusivity.
   The three velocity components are
assumed to satisfy $v_z^2 \ll c_{s0}^2$
and  $v_r^2 \ll v_\phi^2$.
Consequently,  $v_\phi(r,z)$ is  close in
value to the Keplerian value $v_K(r)\equiv (GM/r)^{1/2}$.
Thus, $\partial v_\phi/\partial r = -(1/2)(v_\phi/r)$ 
to a good approximation.

    With these assumptions, the radial component
of equation (1) gives
\begin{equation}
\rho \left({GM \over r^2} -{v_\phi^2 \over r}\right)
= - {\partial p \over \partial r}
+{1\over 4 \pi} B_z {\partial B_r \over \partial z}+F_r^\nu
\end{equation}
     The dominant viscous force is $F_r^\nu=
-\partial T_{rz}^\nu/\partial z$ with $T_{rz}^\nu =
-\rho \nu \partial v_r/\partial z$.

   We normalize the field components by $B_0 = B_z(r,z=0)$,
with  $b_r=B_r/B_0$, $b_\phi = B_\phi/B_0$, 
and $b_z = B_z/B_0\approx 1$.
   Also, we define $u_\phi \equiv v_\phi(r,z)/v_K(r)$ and
the accretion speed $u_r \equiv -~v_r/(\alpha c_{s0})$.
  For the assumed dipole field symmetry, $b_r$ and $b_\phi$
are odd functions of $\zeta$ whereas $u_r$ and $u_\phi$
are even functions.

   Equation (5) then gives
\begin{equation}
{\partial b_r  \over \partial \zeta}
= {\beta \tilde{\rho}\over \varepsilon} ~~
\big(1 - k_p~\varepsilon^2 - u_\phi^2 \big) +\alpha^2\beta
{\partial \over \partial \zeta}
\left(\tilde{\rho} g{\partial u_r \over \partial \zeta} \right)~,
\end{equation}
where $\tilde{\rho} \equiv \rho(r,z)/\rho_0$ with
$\rho_0\equiv \rho(r,z=0)$.
   The midplane plasma beta is 
\begin{equation}
\beta \equiv {4\pi \rho_0c_{s0}^2 \over B_0^2}~,
\end{equation}
where $k_p \equiv - \partial \ln p/\partial \ln r$ is
assumed of order unity and $p=\rho c_s^2$.
   Note that $\beta =c_{s0}^2/v_{A0}^2$, where
$v_{A0}=B_0/(4\pi\rho_0)^{1/2}$ is the midplane
Alfv\'en velocity. 
   The rough condition for the MRI instability
and the associated turbulence in the disk is $\beta >1$
(Balbus \& Hawley 1998). 
   In the following we assume $\beta >1$, which
we refer to as a weak magnetic field.

      The $\phi-$component of equation (1)
gives
\begin{equation}
{\partial b_\phi \over \partial \zeta}
={ \alpha \beta \tilde{\rho}\over 2} ~
\big( 3 \varepsilon k_\nu g - u_r \big)
-{\alpha \beta \over \varepsilon}
{\partial \over \partial \zeta}\left(\tilde{\rho}g
{\partial u_\phi \over \partial \zeta}\right)~,
\end{equation}
where $k_\nu \equiv \partial \ln(\rho c_{s0}^2 r^2/h)/\partial \ln(r)>0$
is of order unity.
   In addition to the well-know  viscous force [$F^\nu_\phi(a) =
-r^{-2}\partial(r^2 T_{r\phi}^\nu)/\partial r$ with
$ T_{r\phi}^\nu=-\rho \nu r\partial(v_\phi/r)/\partial r$] which
gives the term $\propto k_\nu$, we must
include the force contribution
  $F_\phi^\nu(b)= -\partial T_{\phi z}^\nu/\partial z$
with $T_{\phi z}^\nu = -\rho \nu \partial v_\phi/\partial z$. 
This gives the second derivative term in equation (8).

    Note that integration of equation (8) from 
$\zeta=0$ (where $b_\phi=0$ and $\partial u_\phi/\partial \zeta=0$) to $\zeta_S+\epsilon$ (where $g=0$) gives
$$
b_{\phi S+}={1\over 2} \alpha \beta\tilde{\Sigma}
\big(3\varepsilon k_\nu - \overline{u}_r \big)~,
$$  
 where 
$\overline{u}_r\equiv \int_0^{\zeta_S} d\zeta~ \tilde{\rho} u_r/\tilde{\Sigma}$ is the average accretion speed, $\tilde{\Sigma}\equiv \int_0^{\zeta_S} d\zeta~ \tilde{\rho}$, and the $S+$ subscript indicates evaluation at $\zeta=\zeta_S+\epsilon$.
    The average accretion speed, written as
\begin{equation}
\overline{u}_r = u_0 - {2 b_{\phi S+}\over
\alpha \beta \tilde{\Sigma}}~,
\end{equation}
is the sum of a viscous
contribution, $u_0\equiv 3\varepsilon k_\nu$, and a magnetic
contribution ($\propto b_{\phi S+}$) due to the loss of angular
momentum from the surface of the disk where
necessarily $b_{\phi S+} \leq 0$ (Lovelace,
Romanova, \& Newman 1994).   Equation (9)
is discussed further in \S 5.
    The continuity equation implies that
$r h \rho_0\tilde{\Sigma}(\alpha c_{s0} \overline{u}_r )$ is
independent of $r$.

    The $z-$component of equation (1) gives
\begin{equation}
{\partial p \over \partial \zeta}=
-\rho c_{s0}^2\zeta - {\rho_0 c_{s0}^2\over 2 \beta}
{\partial \over \partial \zeta}\big(b_r^2 +b_\phi^2\big)~.
\end{equation}
  The neglected viscous force in this equation,
$ r^{-1}\partial(rT_{rz}^\nu)/\partial r$ with 
$T_{rz}=-\rho \nu \partial v_r/\partial z$, is smaller
than the retained pressure gradient term by a factor of
order $\alpha^2 \varepsilon \ll 1$.
   The term involving the magnetic field describes the
magnetic compression of the disk 
because $b_r^2+b_\phi^2$ at the surface of the disk
is larger than its midplane value which is zero
(Wang, Sulkanen, \& Lovelace  1990).
For $\beta \gg 1$ the compression effect is small and
can be neglected.

     As mentioned we assume $p \propto \rho^\gamma$
which can be written as 
$p=\rho c_{s0}^2 (\rho/\rho_0)^{\gamma-1}$. 
    Thus
\begin{equation}
\tilde{\rho}={\rho \over \rho_0} =\left(
1-{(\gamma-1)\zeta^2 \over 2\gamma}\right)^{1/(\gamma-1)}~,
\end{equation}
for $\beta \gg 1$.   
    The density goes to zero at
$\zeta_m = [2\gamma/(\gamma-1)]^{1/2}$.
   However, before this distance is reached the
MRI turbulence is suppressed, and $g(\zeta)$ in
equation (3) is effectively zero.

The toroidal component of Ohm's law 
(equivalent to equation 2),
$J_\phi= \sigma ({\bf v \times B})_\phi$,
with $\sigma=c^2/(4\pi \eta)$,  gives
\begin{equation}
{\partial b_r \over \partial \zeta} ={{\cal P}\over g} {u_r }~.
\end{equation}
   Multiplying this equation by $g$,  
integrating from $\zeta_S -\epsilon$ to
$\zeta_S +\epsilon$, and using the facts
that    $\partial g/\partial \zeta = -\delta(\zeta-\zeta_S)$
and that $|u_r|$ is bounded implies that $b_{rS+}=
b_{rS-}$.  That is, there is no jump in $b_r$
across the highly conducting surface layer.

    Note that multiplying equation (12)
by $g$ and     integration  from $\zeta =0$
to $\zeta_S +\epsilon$ gives
\begin{equation}
b_{rS} = {\cal P}\zeta_S  \langle u_r \rangle~,
\end{equation}
where $\langle .. \rangle =\int_0^{\zeta_S}d\zeta(..)/\zeta_S$.

The other components of Ohm's law give
\begin{equation}
{\partial u_\phi \over \partial \zeta}
={3\varepsilon \over 2} b_r - 
{\alpha \varepsilon \over {\cal P}}{\partial \over \partial \zeta}
\left( g 
{\partial b_\phi \over \partial \zeta}\right)~.
\end{equation}

   Combining equations (6) and (12) gives
\begin{equation}
u_r ={\beta \tilde{\rho} g \over \varepsilon{\cal P}}
\big(1-k_p\varepsilon^2 - u_\phi^2\big)+{\alpha^2\beta g\over {\cal P}}
{\partial \over \partial \zeta}
\left(\trho g{\partial u_r \over \partial \zeta}\right)~.
\end{equation}
For thin disks, $\varepsilon \ll 1$, and $\beta > 1$,
we have $u_\phi = 1 +\delta u_\phi$ with
$(\delta u_\phi)^2 \ll 1$ which follows from the
integral of equation (6).
  Consequently, 
\begin{equation}
\delta u_\phi = -{k_p\varepsilon^2\over 2}
-{\varepsilon {\cal P}u_r \over 2\beta \tilde{\rho}g}
+{\alpha^2 \varepsilon \over 2 \tilde{\rho}}
{\partial \over \partial \zeta}
\left(\tilde{\rho}g {\partial u_r \over \partial \zeta}\right)~,
\end{equation}
 to a good approximation.
   
   We first take the derivative of equation (14) and then
substitute the $b_r$  derivative  with equation (12).
In turn, the $u_\phi$ derivatives can be put in terms
of $u_r$ and its derivatives using equation(16).
In this way we obtain
\begin{eqnarray}
{\alpha^4\beta^2}
{\partial^2 \over \partial \zeta^2}
\left(g{\partial \over \partial \zeta}
\left(\trho g{\partial \over \partial \zeta}
\left({1\over \trho}{\partial \over \partial \zeta}
\left(\trho g {\partial u_r \over \partial \zeta}\right)\right)
\right)\right)
\nonumber \\
-~\alpha^2\beta {\cal P}
{\partial^2 \over \partial \zeta^2}
\left(g {\partial \over \partial \zeta}
\left(\trho g{\partial \over \partial \zeta}
\left({u_r \over \trho g}\right)\right)\right)
\nonumber\\
-~\alpha^2\beta {\cal P}
{\partial^2 \over \partial \zeta^2}\left({1\over \trho}
{\partial \over \partial \zeta}
\left(\trho g {\partial u_r \over \partial \zeta}\right)\right)
\nonumber \\
+~{\alpha^2\beta^2 }
{\partial^2 \over \partial \zeta^2}\bigg(\trho g \big(u_r- gu_0
\big)\bigg) 
+{\cal P}^2 {\partial^2 \over \partial \zeta^2}\bigg({u_r \over \trho g}
\bigg)
\nonumber \\
+~3\beta{\cal P }^2{u_r \over g}=0~.
\end{eqnarray}
  The equation can be integrated from $\zeta =0$
out to the surface of the disk $\zeta_S$ where 
boundary conditions apply.
  Because $u_r$ is an even function of $\zeta$, the
odd derivatives of $u_r$ are zero at $\zeta=0$
and one needs to specify $u_r(0)$, $u^{\prime \prime}_r(0)$,
and $u^{iv}_r(0)$.  
   A  ``shooting method'' can be
applied where the values of
$u_r(0)$, $u_r^{\prime \prime}(0)$, and $u_r^{iv}(0)$
are adjusted to satisfy the boundary conditions.
   Once $u_r(\zeta)$ is calculated,   equations (8), (12),
and (14) can be integrated to obtain $b_\phi(\zeta)$,
$b_r(\zeta)$, and $u_\phi(\zeta)$.

     For specificity we take
\begin{equation}
g(\zeta)=\left(1-{\zeta^2\over \zeta_S^2}\right)^\delta ~,
\end{equation}
where
 $\zeta_S  <\zeta_m$ and $\delta \ll 1$.  
    That is, we neglect the ratio of the Spitzer
diffusivity on the surface of the disk  to
its value in the central part of the disk.   
  An  estimate of $\zeta_S$ can be made
by noting that $\beta(\zeta)=4\pi p(\zeta)/B_0^2 =
\beta (\tilde{\rho})^\gamma \approx 1$
at $\zeta_S$.  
    This gives $\zeta_S^2/\zeta_m^2=
1-\beta^{-(\gamma-1)/\gamma}$ and 
$\rho_S/\rho_0 = \beta^{-1/\gamma}$.

    If ${\cal E}$ denotes the fraction of the
disk mass accretion rate which goes into outflows,
then    we can estimate the vertical speed of 
matter at the disk's surface as $v_z(\zeta_S) \sim
{\cal E} (2h/r) (\rho_0/\rho_S)|\overline{v}_r|
={\cal E}(2h/r) \beta^{1/\gamma}|\overline{v}_r|$.
Our neglect of $v_z$ evidently requires that
${\cal E}(2h/r) \beta^{1/\gamma} \ll 1$.

\section{Boundary Conditions}

   We restrict our attention to physical solutions
which (a) have  net mass accretion,
\begin{equation}
\dot{M}=4\pi r h \rho_0 \alpha c_{s0}\tilde{\Sigma}~
\overline{u}_r > 0~,
\end{equation}
and (b) have $b_{\phi }\leq 0$ on the disk's surface.
   This condition on $b_{\phi S+}$ corresponds to an
efflux of angular momentum and energy
(or their absence) from
the disk to its corona rather than the reverse.
  From equation (9), the condition $b_{\phi S+} \leq 0$
 is the same as $\overline{u}_r \geq  u_0$,
 where $u_0$ is the minimum (viscous) accretion speed.
   Note that $\overline{u}_r/u_0 -1$ is the fraction of the
accretion power which goes into the outflows or jets (Lovelace
et al. 1994).   
    Clearly, the condition on $b_{\phi S+}$ implies that $\dot{M}>0$
so that there is  only one condition.    

    In general there is a continuum of values of $b_{\phi S+} \leq 0$
for the considered solutions inside the disk.
   The value of $b_{\phi S +}$ can be determined by matching
the calculated fields $b_{rS}$ and $b_{\phi S+}$ onto
an external field and flow solution as discussed in \S 5.
 
    From equation (12) we found that there is no
jump in $b_r$ across the conducting surface layer.
  Thus, integration of equation (6) from 
$\zeta_S-\epsilon$ to $\zeta_S+\epsilon$  implies that
\begin{equation}
{\partial u_r\over \partial \zeta}\bigg|_{\zeta_S-} =0~.
\end{equation}
This represents a second condition on the disk solutions.

   Integration of equation (14) from
$\zeta_S-\epsilon$ to $\zeta_S+\epsilon$ gives
$$
u_{\phi S+} -u_{\phi S -} = 
{\alpha \varepsilon \over {\cal P}} 
{\partial b_\phi \over \partial \zeta}\bigg|_{\zeta_{S-}}~. 
$$
  This velocity jump must be zero as can be shown by
inspection of the total angular momentum flux-density
in the $z-$direction,   $r T_{\phi z} = -B_\phi B_z/4\pi 
-\rho \nu( \partial v_\phi/\partial z)$, where the first term
is the magnetic stress and the second is the viscous 
stress.   A jump in $v_\phi$ would give a delta-function
contribution to the viscous stress which cannot be
balanced by the magnetic stress.
      Therefore
\begin{equation}
{\partial b_\phi \over \partial \zeta} \bigg{|}_{\zeta_{S-}} =0~.
\end{equation}
   This  is a third condition on the disk solutions.

    Integration of equation (8) from
$\zeta_S-\epsilon$ to $\zeta_S+\epsilon$  gives
\begin{equation}
b_{\phi S+} -b_{\phi S -} = 
{\alpha \beta \tilde{\rho}_S \over \varepsilon} 
{\partial u_\phi \over \partial \zeta}\bigg|_{\zeta_{S-}}~.
\end{equation}
  This equation is equivalent to the continuity
of the angular momentum flux-density across the surface
layer, namely, $rT_{ \phi z}^+ =-B_\phi^+B_z/4\pi$ (above the
layer)
is equal to $rT_{\phi z}^- = -B_\phi^- B_z/4\pi -
\rho \nu (\partial v_\phi /\partial z)|_-$ (below the layer).
  The jump $b_{\phi S+} -b_{\phi S -}$ corresponds
to a radial surface current flow in the highly conducting
surface layer of the disk, ${\cal J}_r =
-(c/4\pi)B_0(b_{\phi S+} -b_{\phi S -})$.

\begin{inlinefigure}
\centerline{\epsfig{file=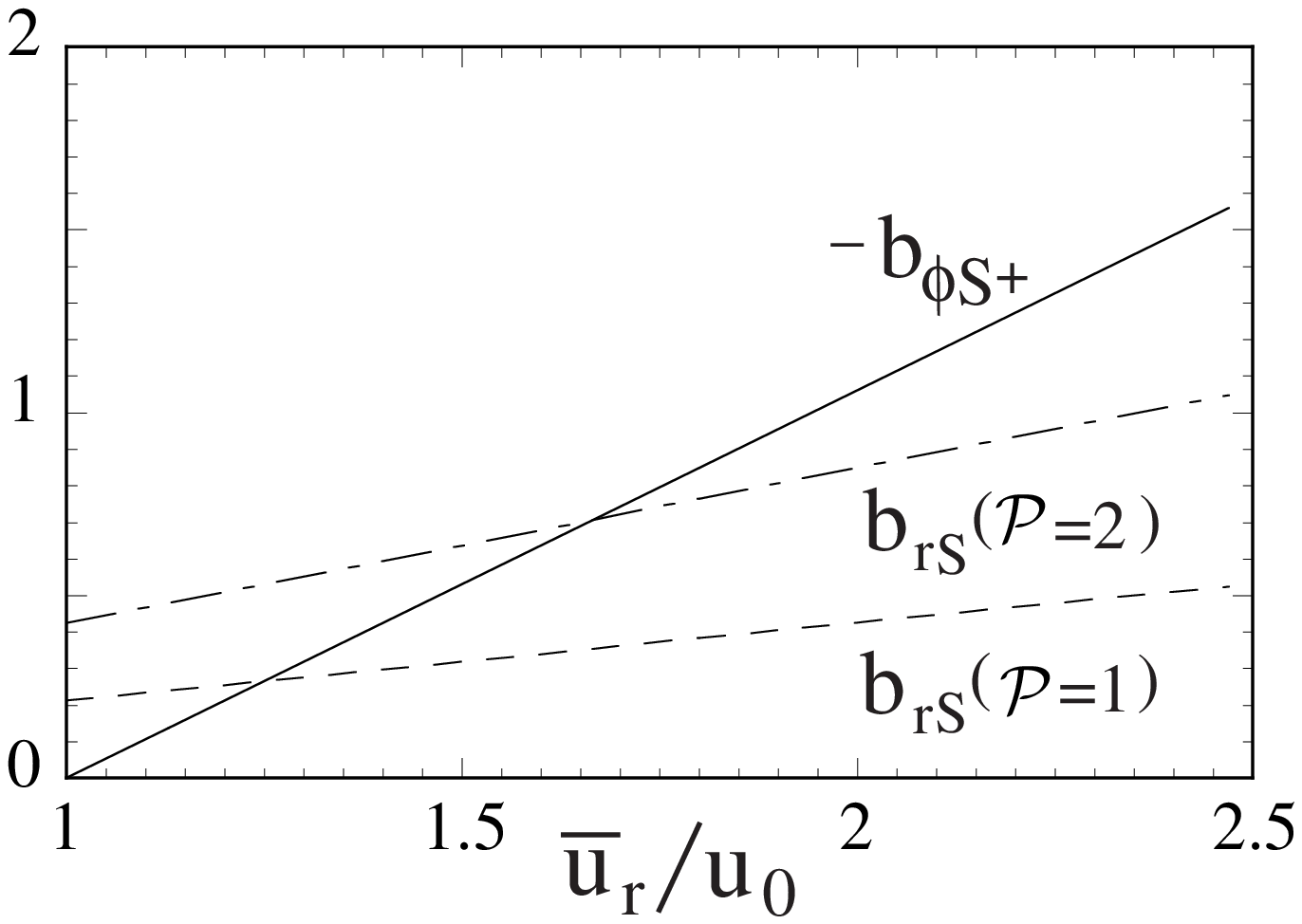,
height=2.2in,width=3.5in}}
\epsscale{0.8}
\figcaption{Radial and toroidal field components
(normalized to $B_z$) at the disk's surface
as a function of the average accretion speed
$\overline{u}_r$ (normalized by the
viscous accretion speed $u_0$).
   For this plot 
$\beta=100$ and Prandtl
numbers ${\cal P}=1$ and $2$.  Note that
$b_{\phi S+}$ is given by equation (10) and is independent
of ${\cal P}$ and $b_{rS}$ is given by equation (13).}
\end{inlinefigure}

 \section{Internal Solutions}

    Here, to simplify the analysis we
consider the limit where 
$\gamma \rightarrow  \infty$ in equation
(11) and $\delta \rightarrow 0$ in
equation (18).    
   Then, $\zeta_S \rightarrow \zeta_m$ and
both $\tilde{\rho}$ and $g$
are unit step functions going to zero at $\zeta_m =\sqrt{2}$.
Also,  $\overline{u}_r = \langle u_r \rangle$ and $\tilde{\Sigma}=
\sqrt{2}$.
   Thus the above physical condition 
$\overline{u}_r \geq 3 \varepsilon k_\nu=u_0$ implies
that $b_{rS} \geq u_0 \zeta_S{\cal P}$
from equation (13).   We assume $k_p$ and $k_\nu =1$.

   The solutions to equation (17) are
$u_r \propto \exp(ik_j \zeta)$ (with $j=1,~2,~3$), where
\begin{equation}
 \alpha^4\beta^2(k_j^2)^3+2{\cal P}\alpha^2\beta(k_j^2)^2+
 (\alpha^2 \beta^2+{\cal P}^2)k_j^2-3\beta{\cal P}^2  =0~,
\end{equation}
is a cubic in $k_j^2$.      
  The discriminant of the cubic is negative so that there
is one real root, $k_1^2$, and a complex conjugate
pair of roots,
$k_{2,3}^2$.    
    Because $u_r$ is an even function of $\zeta$ we
can write
\begin{eqnarray}
u_r& = &
a_1 \cos(k_0\zeta)+a_2 \cos(k_r\zeta) \cosh(k_i\zeta)
\nonumber\\
&+&a_3 \sin(k_r\zeta)\sinh(k_i\zeta)~,
\end{eqnarray}
where $k_0 =\sqrt{k_1^2}$, 
$k_r ={\rm Re}(\sqrt{k_2^2})$, and 
$k_i ={\rm Im}(\sqrt{k_2^2})$.

    The three unknown constants
$a_1, a_2, a_3$ are reduced to two
by imposing equation (20).   The
two are then reduced to one constant
by imposing equation (21).
   The remaining constant is restricted
 by the condition $b_{\phi S+} \leq 0$.

      We consider a thin disk,   
$\varepsilon =h/r=0.05$,  and  a
viscosity parameter $\alpha=0.1$.  
     Figure 1 shows the dependences of
the surface field components on the
average accretions speed for $\beta=100$
and two values of ${\cal P}$. 
The $b_{\phi S}$ dependence is given by
equation (9) and is independent of ${\cal P}$
while the $b_{rS}$ is given by equation (13).

\begin{inlinefigure}
\centerline{\epsfig{file=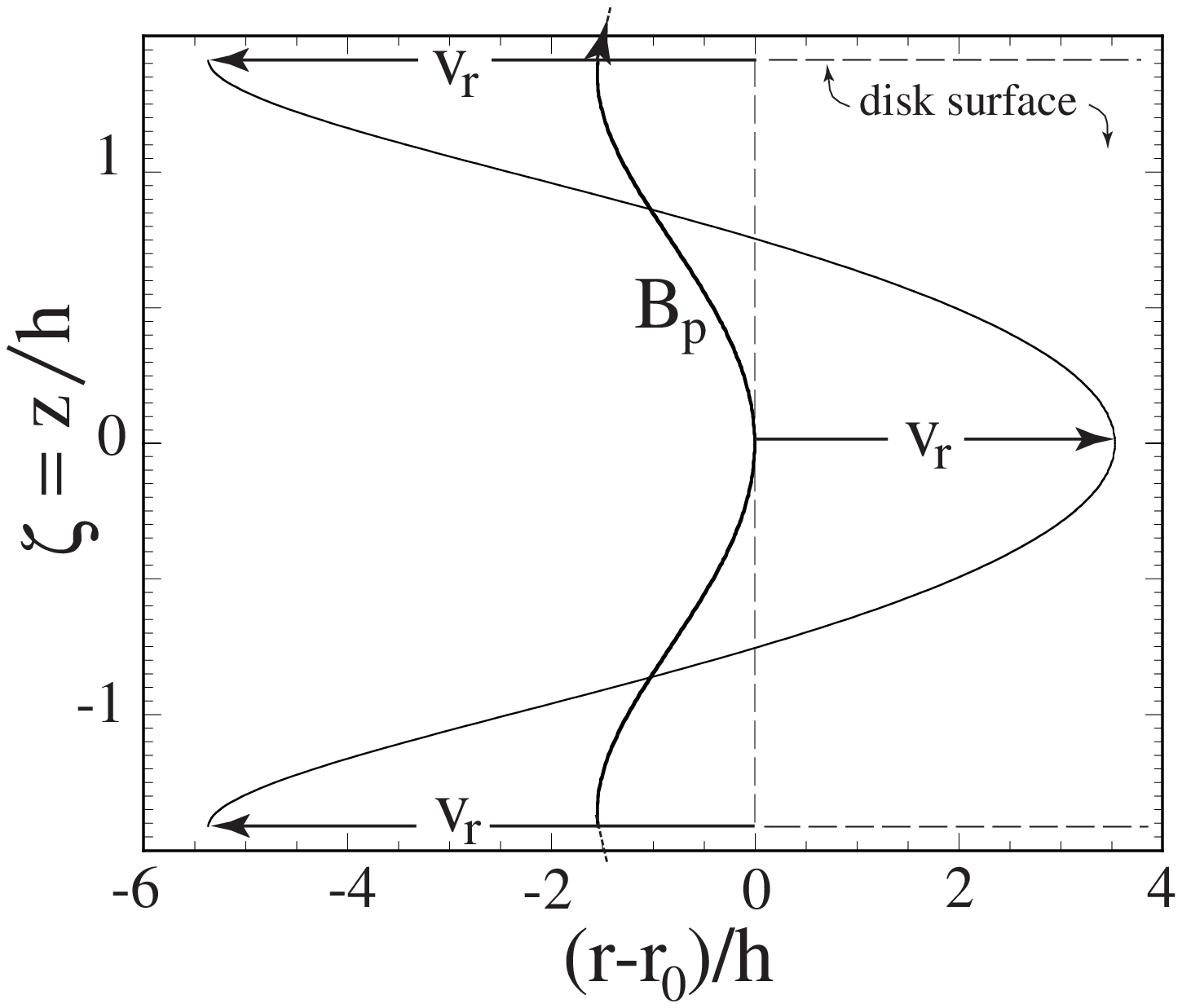,
height=3.2in,width=3.5in}}
\epsscale{0.8}
\figcaption{
Radial flow speed $v_r = - u_r$ (normalized
to $\alpha c_{s0}$) as a function
of $\zeta=z/h$ and a sample poloidal $(B_r,B_z)$
magnetic field line
for $\beta=10^2$ and ${\cal P}=1$.  Also, $r_0$ is
a reference radius.
}
\end{inlinefigure}

    Figure 2 shows both the profile of the
accretion speed $u_r(\zeta)$ and
the shape of the poloidal magnetic field ${\bf B}_p$
for $\beta=100$ and ${\cal P}=1$.  
        Figure 3 shows the profiles of the toroidal
magnetic field $b_\phi$ and the fractional deviation of the
toroidal velocity from the Keplerian value,
$\delta u_\phi =(v_\phi - v_K)/v_K$.
    For this case, $\overline{u}_r/u_0 =1.30$,
$b_{\phi S +} =-0.321$, and $b_{rS}=0.276$.    
    In this case (and a range of others discussed below),
we find that there is radial inflow of the top and
bottom parts of the disk whereas there is radial {\it outflow}
$u_r <0$ of the part of the disk around the midplane.
    Inspection of the flow/field solution
shows that the top and bottom parts of the disk 
lose angular momentum (by the vertical angular momentum
flux $rT_{\phi z}$)
 {\it both} to (a) magnetic
winds or jets from the disk's surfaces {\it and}  to (b) the
vertical flow of angular momentum to the midplane
part of the disk which  flows radially outward.

    We find that the
flow pattern is the same as in Figure 2 for $10 \leq \beta \leq 200$
and ${\cal P} \geq 1$.
    For larger values of $\beta$  and ${\cal P}=1$, the
flow pattern changes from that in Figure 2 to that  in
Figure 4 for $\beta =300$.  
  However, for $\beta =300$
and ${\cal P}=2$, the flow pattern is again similar to
that in Figure 2.     For $\beta = 10^3$ and ${\cal P}=3$
the flow pattern is also the same as in Figure 2.
    For smaller viscosity values, $\alpha \leq 0.03$,
and $\beta =100$ there are multiple channels with for
example three layers of radial inflow and two layers of
radial outflow in the disk.

\begin{inlinefigure}
\centerline{\epsfig{file=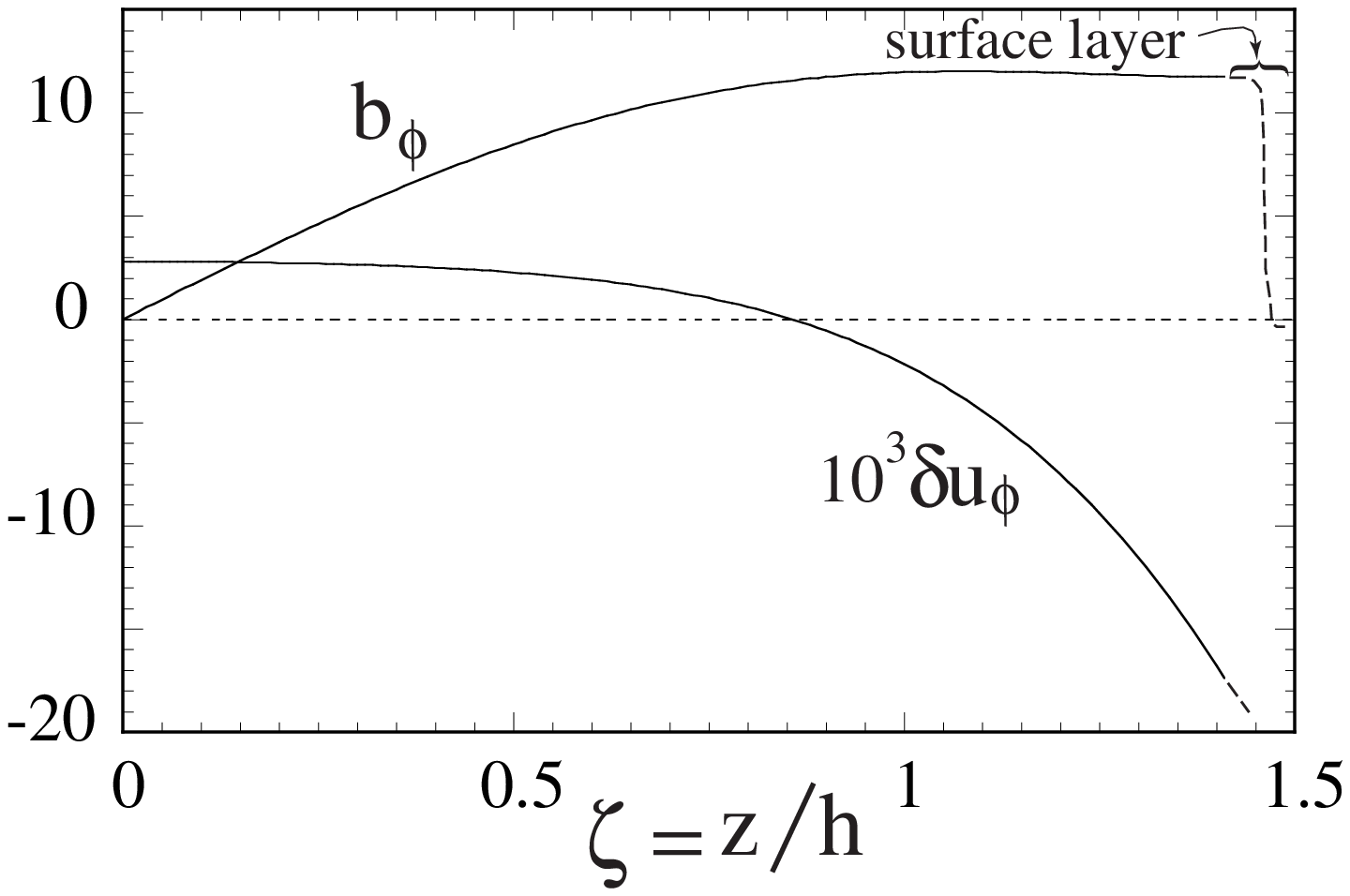,
height=2.2in,width=3.5in}}
\epsscale{0.8}
\figcaption{Toroidal magnetic
field $b_\phi =B_\phi/B_z$ and
toroidal velocity $\delta u_\phi =(v_\phi - v_K)/v_K$
(with $v_K$ the Keplerian velocity)
 for the case where
$\beta=100$ and ${\cal P}=1$.  
The jump in the toroidal magnetic field at the
disk's surface is shown by the dashed line.}
\end{inlinefigure}

    Figure 4 shows the profile of the
accretion speed $u_r(\zeta)$ and
the shape of the poloidal magnetic field ${\bf B}_p$
for $\beta=300$ and ${\cal P}=1$.  
        Figure 5 shows the profiles of the toroidal
magnetic field $b_\phi$ and the fractional deviation of the
toroidal velocity from the Keplerian value,
$\delta u_\phi =(v_\phi - v_K)/v_K$.
    For this case, $\overline{u}_r/u_0 =1.5$,
$b_{\phi S +} =-1.59$, and $b_{rS}=0.318$.    
     Inspection of the flow/field solution
shows that the angular momentum
flux $rT_{\phi z}>0$  in the top half of the disk
and at the disk's surface this flux goes into
an outflow or jet.

\begin{inlinefigure}
\centerline{\epsfig{file=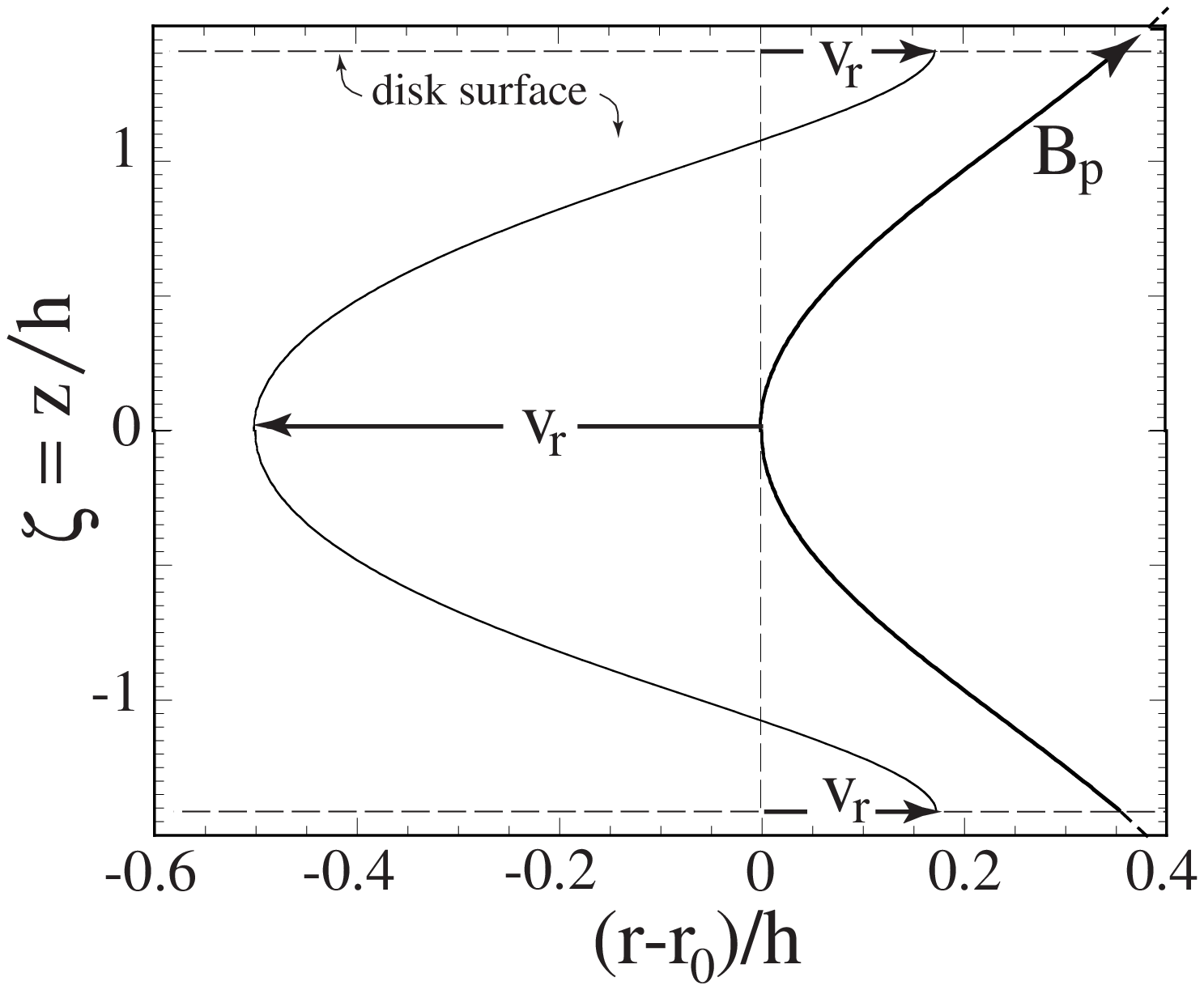,
height=3.2in,width=3.5in}}
\epsscale{0.8}
\figcaption{
Radial flow speed $v_r=-u_r$ (normalized
to $\alpha c_{s0}$) as a function
of $\zeta=z/h$ and a sample poloidal $(B_r,B_z)$
magnetic field line
for $\beta=300$ and ${\cal P}=1$.  Also,
$r_0$ is a reference radius.}
\end{inlinefigure}

\section{External Solutions}

   As mentioned in \S 3, the value of $b_{\phi S+}\leq 0$ is not
fixed by the solution for the field and flow inside the disk.
     Its value can be determined by matching the calculated
surface fields $b_{r S}$  and $b_{\phi S+}$ onto an external
magnetic wind or jet solution.   Stability of the wind or jet
solution to current driven kinking is predicted to
limit the ratio of the toroidal to axial magnetic field
components at the disk's surface $|b_{\phi S +}|$ to values
 $\lesssim {\cal O}(2\pi r/L_z)$ (Hsu \& Bellan 2002; Nakamura, Li,
 \& Li 2007),
 where $L_z$ is the length-scale of  field divergence
 of the wind or jet at the disk surface.  
      From known wind and jet solutions we
estimate $2\pi r/L_z \approx \pi$ (Lovelace,
Berk, \& Contopoulos 1991; Ustyugova et al. 1999; 
Ustyugova et al. 2000;  Lovelace et al. 2002).      
      Recall that $\overline{u}_r/u_0 -1= 2|b_{\phi S +}|/
(\alpha \beta \tilde{\Sigma} u_0)$ (from equation 9) is
the faction of the accretion power going into the jets
or winds.   
     For the mentioned upper limit on $|b_{\phi S+}|$,
we find      $\overline{u}_r/u_0 -1 \lesssim {\cal O}[2\pi/(\alpha \beta
 \tilde{\Sigma} u_0)]$.    
      From equation (13) we have $b_{rS} =
 ({\cal P}\zeta_S u_0)(\langle{u}_r\rangle/u_0)$.
     Therefore, for $\beta \gg 1$ and
$\langle{u}_r\rangle \approx u_0$, we have
 $ b_{rS} \approx {\cal P}\zeta_S u_0$.

   The matching of internal and external field/flow solutions has been
carried out by K\"onigl (1989) and Li (1995) for the case of
self-similar  [$B_z(r,0) \sim r^{-5/4}$]
magnetocentrifugally outflows  from the disk's surface.    
     These outflows occur under conditions where the
poloidal field lines at the disk's surface are tipped relative to the
rotation axis by more than $30^\circ$  which 
corresponds to $b_{rS} > 3^{-1/2} \approx 0.577$
(Blandford \& Payne 1982).    
     The outflows typically  carry a significant mass flux.     
   For  the internal field/flow solutions
discussed in \S 4 with $\beta \gg1$, we conclude that $b_{rS}$ is 
sufficiently large for magnetocentrifugal outflows only
for turbulent magnetic Prandtl numbers, ${\cal P} \gtrsim 2.7$.
    Shu and collaborators (e.g., Cai et al. 2008, and references therein)
have developed detailed `X-wind' models which depend on the
disk having Prandtl numbers  larger than
unity.    
      Recent MHD simulations by Romanova et al. (2008) 
provides evidence of  conical or X-wind type  outflows for 
Prandtl numbers $\geq1$.

\begin{inlinefigure}
\centerline{\epsfig{file=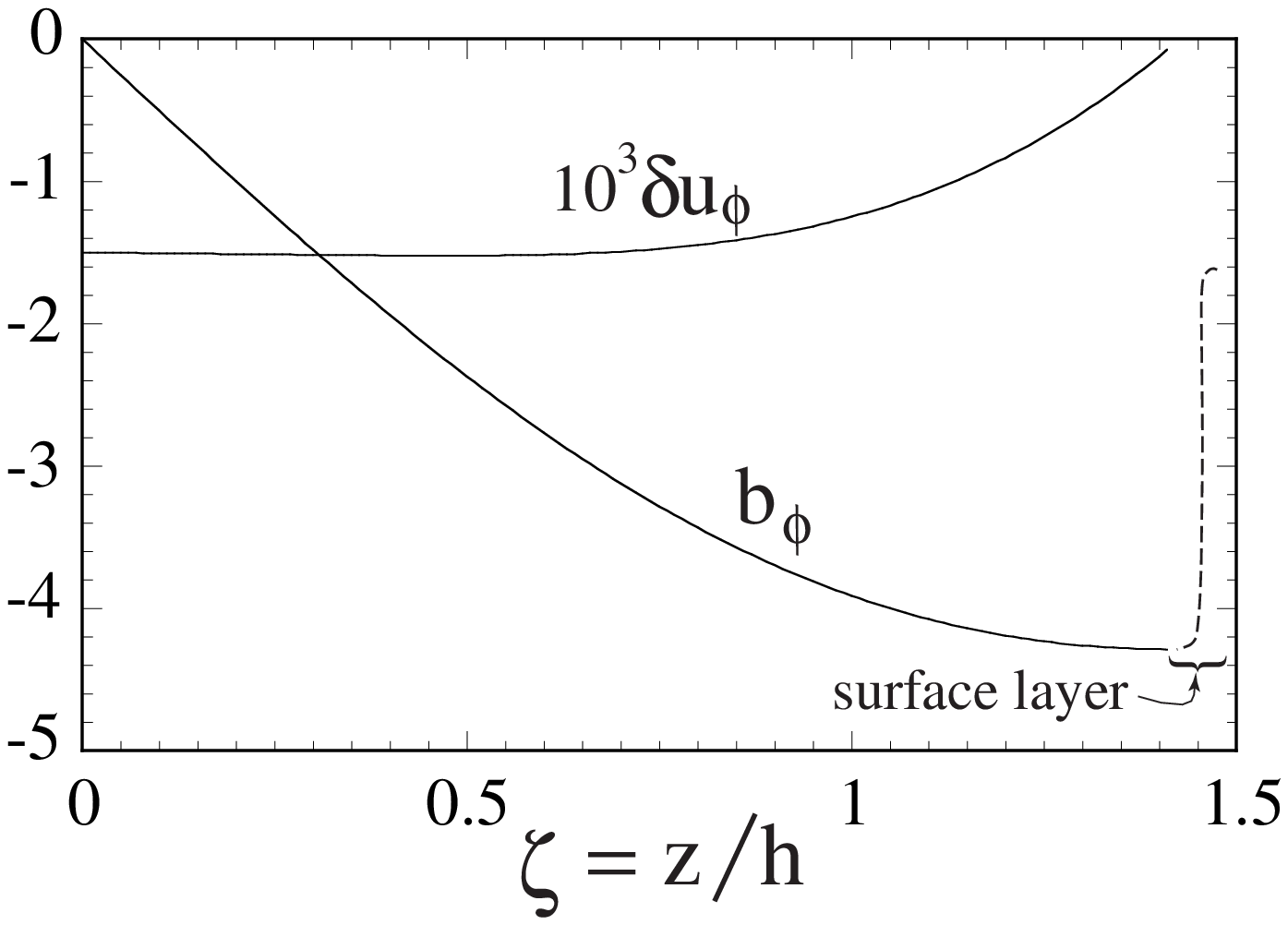,
height=2.2in,width=3.5in}}
\epsscale{0.8}
\figcaption{Toroidal magnetic
field $b_\phi =B_\phi/B_z$ and
toroidal velocity $\delta u_\phi =(v_\phi - v_K)/v_K$
(with $v_K$ the Keplerian velocity)
 for the case where
$\beta=300$ and ${\cal P}=1$.  
The jump in the toroidal magnetic field at the
disk's surface is shown by the dashed line.}
\end{inlinefigure}

      For Prandtl numbers say ${\cal P} \lesssim 2.7$, the values of
$b_{rS}$ are too small for there to be a magnetocentrifugal
outflow.   
    In this case there is  an outflow of 
electromagnetic energy and angular momentum from the disk
(with little mass outflow) in the
form of a magnetically
dominated or `Poynting flux jet'  (Lovelace, Wang, \& Sulkanen 1987;
Lovelace, et al. 2002)
also referred to as a `magnetic tower jet' (Lynden-Bell 1996, 2003).
   MHD simulations have established the  occurrence of
Poynting-flux jets under different conditions (Ustyugova et al. 2000,
2006;   Kato, Kudoh, \& Shibata 2002; Kato 2007).
   Laboratory experiments  have
allowed the generation of magnetically dominated   jets
(Hsu \& Bellan 2002;  Lebedev et al. 2005).

\section{Discussion}

      A study is made of
stationary axisymmetric accretion flows  $[v_r(z),v_\phi(z),v_z=0]$
 and the  large-scale, weak
magnetic field [$B_r(z),B_\phi(z), B_z\approx{\rm const}$] taking into
account the turbulent viscosity and diffusivity
due to the MRI and the fact that the turbulence 
vanishes at the surface of the disk  as
discussed by Bisnovayi-Kogan
\& Lovelace (2007) and  Rothestein \& Lovelace (2008).
   We derive a sixth-order differential equation
for the radial flow velocity $v_r(z)$ (equation 17) which 
depends mainly on the ratio of the midplane
thermal to magnetic pressures $\beta >1$ and
the Prandtl number of the turbulence
${\cal P}=$ viscosity/diffusivity. 
     Boundary conditions at the disk's
surfaces take into account the outflow of
angular momentum to magnetic
winds or jets and allow for  current
flow in the highly conducting surface layers.  
  In general we find that there is a radial
surface current but  no toroidal surface current.
  The stability of this surface current layer is
unknown and remains to be studied.  
If the layer is  unstable to kinking this may 
cause a thickening of the current layer. 
    We argue that stability of the winds or jets will
limit  the ratio of the
toroidal to axial field at the disk's surface $|b_{\phi S+}|$ to values
$\lesssim \pi$.
   The stationary solutions we find indicate that a
weak ($\beta \gg 1$), large-scale field  does not diffuse away as suggested
by earlier work (e.g., Lubow et al. 1994) which
assumed $b_{rS} \geq 3^{-1/2}$.

     For a wide range of parameters $\beta > 1$ and ${\cal  P}\geq 1$,
we find stationary {\it channel} type flows where 
the  flow is radially {\it outward} near the midplane
of the disk and  radially {\it inward} in the top and bottom parts
of the disk.       Solutions with inward flow near the midplane
and outflow in the top and bottom parts of the disk are also found.
    The solutions with radial outflow near the midplane are of
interest for the outward transport of  chondrules   in protostellar
disks from distances close to the star ($\sim 0.05$ AU) (where they
are melted and bombarded by high energy particles)
to larger distances ($> 1$ AU) where they are 
observed in the Solar system.  Outward transport of chondrules
from distances $\sim 0.05$ AU to $>1$ AU
by an X-wind has been discussed by Shu et al. (2001).

     The flow/field solutions found here in a 
viscous/diffusive disk and are different from    the  
exponentially growing channel flow solutions
found by Goodman and Xu (1994) for an MRI in
an ideal MHD unstable shearing box.    Channel
solutions in viscous/diffusive disks were found
earlier by Ogilvie \& Livio (2001) and by Salmeron,
K\"onigl, \& Wardle (2007) for conditions different
from those considered here.   
   In general we find that the magnitude of the toroidal
magnetic field component inside the disk is
much larger than the other field components.    
The fact that the viscous accretion speed is very small,
$\sim \alpha \varepsilon c_{s0}$, means that even a
small large-scale field can significantly influence
the accretion flow.
    We find that  Prandtl numbers larger than a critical value
estimated to be $2.7$ are needed in order for there 
to be magnetocentrifugal outflows from
the disk's surface.  For smaller ${\cal P}$, electromagnetic outflows are
predicted.  
     Owing to the stability condition, $|b_{\phi S+}| \lesssim \pi$, the 
fraction of the accretion power going into magnetic outflows
or jets is $\lesssim {\rm const}\beta^{-1} \sim B_z^2$.

       Analysis of the time-dependent accretion of the large-scale
${\bf B}-$field is clearly needed to study the amplification
of the field and build up of magnetic flux in the inner region
of the disk.   One method is to use global 3D MHD simulations 
(Igumenshchev et al. 2003; Igumenshchev 2008), but this has
the difficulty of resolving the very thin highly conducting 
surface layers of the disk.   Another method is to generalize
the approach of Lovelace et al. (1994) taking into account
the results of the present work.    This is possible because
the radial accretion time ($r/|\overline{u}_r|$) is typically
much longer than the  viscous diffusion time across the disk ($h^2/\nu$).

We thank an anonymous referee for valuable comments.
The work of G.S.B.-K. was partially supported by RFBR grants
08-02-00491 and 08-02-90106, RAN Program ``Formation
and evolution of stars and galaxies.''
D.M.R. was supported by
an NSF Astronomy and Astrophysics Postdoctoral Fellowship
under award AST-0602259.
R.V.E.L. was supported in
part by NASA grant NNX08AH25G and by
NSF grants AST-0607135 and AST-0807129.


\begin{references}

\reference{} Balbus, S.A., \& Hawley, J.F. 1991, ApJ, 376, 214

\reference{} Balbus, S.A., \& Hawley, J.F. 1998, Rev. Mod. Phys., 70, 1

\reference{} Bisnovatyi-Kogan, G. S., \& Ruzmaikin, A. A. 1974, Ap\&SS, 28, 45 --- 1976, Ap\&SS, 42, 401

\reference{} Bisnovatyi-Kogan, G.S., \& Lovelace, R.V.E. 2007, ApJ, 667, L167



\reference{} Blandford, R.D., \& Payne, D.G. 1982, MNRAS, 199, 883

\reference{} Cai, M.J., Shang, H., Lin, H.H., \& Shu, F.H. 2008,
ApJ, 672, 489

\reference{} De Villiers, J.-P., Hawley, J. F., Krolik, J. H., \& Hirose, S. 2005, ApJ, 620, 878

\reference{} Goodman, J., \& Xu, G. 1994, ApJ, 432, 213

\reference{} Hawley, J. F., \& Krolik, J. H. 2006, ApJ, 641, 103

\reference{} Hirose, S., Krolik, J. H., De Villiers, J.-P., \& Hawley, J. F. 2004, ApJ, 606, 1083

\reference{} Hsu, S.C., \& Bellan, P.M. 2002, MNRAS, 334, 257

\reference{} Igumenshchev, I. V., Narayan, R., \& Abramowicz, M. A. 2003, ApJ, 592, 1042

\reference{} Igumenshchev, I.V. 2008, ApJ, 677, 317

\reference{} Kato, S.X., Kudoh, T., \& Shibata, K. 2002, ApJ, 565, 1035

\reference{} Kato, Y. 2007, A\&SS, 307, 11

\reference{} K\"onigl, A. 1989, ApJ, 342, 208

\reference{} Lebedev, S.V., Ciardi, A., Ampleford, D.J., Bland, S.N.,
Bott, S.C., Chittenden, J.P., Hall, G.N., Rapley, J., Jennings, C.A., Frank, A.,
Blackman, E.G., \& Lery, T. 2005, MNRAS, 361, 97

\reference{} Li, Z.-Y., 1995, ApJ, 444, 848

\reference{} Lovelace, R.V.E. 1976, Nature, 262, 649

\reference{} Lovelace, R.V.E., Wang, J.C.L., \& Sulkanen, M.E. 1987,
  ApJ, 315, 504
  
\reference{} Lovelace, R.V.E., Berk, H.L., \& Contopoulos, J. 1991,
ApJ, 379, 696  
  



\reference{} Lovelace, R.V.E., Romanova, M.M., \& Newman, W.I. 1994,
ApJ, 437, 136 --- 1997, ApJ, 484, 628

\reference{} Lovelace, R. V. E.,  Li, H.,  Koldoba, A. V., 
Ustyugova, G. V., \& Romanova, M. M. 2002, ApJ, 572, 445



\reference{} Lubow, S. H., Papaloizou, J. C. B., \& Pringle, J. E. 1994, MNRAS, 267, 235

\reference{} Lynden-Bell, D. 1996, MNRAS, 279, 389

\reference{} Lynden-Bell, D. 2003, MNRAS, 341, 1360

\reference{} McKinney, J. C., \& Narayan, R. 2007, MNRAS, 375, 513

\reference{} Nakamura, M., Li, H., \& Li, S. 2007, ApJ, 656, 721

\reference{} Narayan, R., Igumenshchev, I.V., \& Abramowicz, M.A. 2003,
PASJ, 55, L69


\reference{} Ogilvie, G.I., \& Livio, M. 2001, ApJ, 553, 158

\reference{} Pessah, M.E., Chan, C.-k, \& Psaltis, D. 2008, MNRAS, 383, 683

\reference{} Romanova, M.M., et al. 2008, in preparation

\reference{} Rothstein, D.M., \& Lovelace, R.V.E. 2008, ApJ, 677, 1221

\reference{} Salmeron, R., K\"onigl, \& Wardle, M. 2007, MNRAS, 375, 177

\reference{} Shakura, N.I., \& Sunyaev, R.A. 1973, A\&A, 24, 337

\reference{} Shu, F.H., Shang, H., Gounelle, M., Glassgold, A.E., \& Lee, T.
2001, ApJ, 548, 1029

\reference{} Spruit, H.C., \& Uzdensky, D.A. 2005, ApJ, 629, 960

\reference{} Ustyugova, G. V., Koldoba, A. V., Romanova, M. M., Chechetkin, V. M.;,
\& Lovelace, R. V. E. 1999, ApJ, 516, 221

\reference{} 	Ustyugova, G. V., Lovelace, R. V. E., Romanova, M. M., Li, H., \& Colgate, S. A. 2000, ApJ, 541, L21

\reference{} Ustyugova, G. V., Koldoba, A. V.,  Romanova, M. M., \&  Lovelace, R. V. E.
2006, ApJ, 646, 304

\reference{} van Ballegooijen, A. A. 1989, in Accretion Disks and Magnetic Fields in Astrophysics, ed. G. Belvedere ( Dordrecht: Kluwer), 99

\reference{} Wang, J.C.L., Sulkanen, M.E., \& Lovelace, R.V.E. 1990,
ApJ, 355, 38






\end{references}
\end{document}